\documentclass[a4paper,10pt]{article}

\usepackage[english]{babel}
\usepackage[utf8]{inputenc}
\usepackage{indentfirst}
\usepackage{amsmath}
\usepackage{amsfonts}
\usepackage{bm}
\usepackage{graphicx}
\usepackage{authblk}
\usepackage{braket}
\usepackage[left=2cm,right=2cm,top=3cm,bottom=3cm]{geometry}
\usepackage[autostyle, english = american]{csquotes}

\MakeOuterQuote{"}

\title{\bf Lateral disorder in Langmuir monolayers:
        theoretical derivations and grazing-incidence
        X-ray diffraction}

\author[1,2]{L.R.~Muftakhova}
\author[2,1,*]{K.V.~Nikolaev}
\author[2]{A.V.~Rogachev}
\author[2]{N.N.~Novikova}
\author[2,3]{B.I.~Ostrovskii}
\author[2]{S.N.~Yakunin}

\affil[1]{Moscow Institute of Physics and Technology, Dolgoprudny, Russia}
\affil[2]{National Research Center Kurchatov Institute, Moscow, Russia}
\affil[3]{Institute of Solid State Physics RAS, Chernogolovka, Russia}
\affil[*]{Correspondence e-mail: \texttt{k.nikolaev@protonmail.com}  }

\date{September 24, 2025}

\begin{document}
\maketitle

\begin{abstract}

    Recent studies of the self-assembly of Langmuir monolayers have revealed novel forms of lateral molecular ordering.
    Such studies typically involve the use of grazing-incidence synchrotron radiation scattering, 
    and the lateral order manifests itself as distinct diffraction patterns. 
    The more intricate the molecular organization, 
    the more complicated the corresponding diffraction pattern is. 
    To the point where standard analysis, i.e., 
    identifying peak positions and solving the crystal structure,
    is insufficient to describe the system. 
    In such cases, 
    a physics-based simulation of the diffraction is required. 
    In this article, 
    we present a versatile theoretical framework for simulating complex structural molecular ordering in Langmuir monolayers. 
    We begin by applying the formalism to a simple case of solid-state monolayers and extend the analysis to describe the structural organization in the collapsed state. 
    The applicability of the method is validated through comparison with experimental data collected at the bending magnet synchrotron beamline.
    
\end{abstract}

\section{Introduction}

\begin{figure}[t!]
        \centering   
        \includegraphics[width=0.5\textwidth]{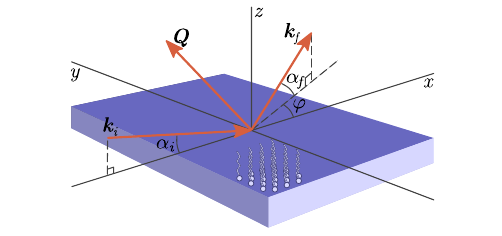}
        \caption{
            The experimental setup for grazing-incidence diffraction (GID). 
            X-ray radiation is directed onto an interface at a small angle, $\alpha_i$, 
            close to the critical angle of total external reflection. 
            The scattering vector, $\bm{Q} = (\bm{Q}_\parallel, Q_z)$, 
            where $\bm{Q}_\parallel$ and $Q_z$ are the horizontal and vertical components of the scattering vector, respectively, 
            it is determined by the incident angle, $\alpha_i$, the exit angle, $\alpha_f$, 
            and the azimuthal angle, $\varphi$.
            The scattered photons are detected above the surface in various directions, 
            corresponding to different values of the vector $\bm{Q}$.
        }
        \label{fig:gid_scheme}
    \end{figure}

    The structure of Langmuir monolayers has been extensively investigated~\cite{kaganer1999structure}.
    Nevertheless, they continue to attract considerable research interest due to their complex phase behavior.
    Recent studies~\cite{singh2025} have once again highlighted intriguing structural reorganization phenomena related to the interaction of Langmuir monolayers with rare earth elements.
    In particular, certain lanthanides induce the formation of a laterally ordered structures that may be commensurate with the molecular ordering within the monolayer.
    The crystalline ordering of Langmuir monolayers is typically probed using X-ray diffraction.
    An experimental configuration suitable for such studies was introduced in~\cite{kjaer1987ordering, dutta1987x}.
    The main idea is to excite an evanescent wave at grazing incidence close to the condition of total external reflection 
    (see~\cite{andreev1985x} for a detailed theoretical review), 
    and to measure its out-of-plane diffraction.
    The experiment geometry is sketched out in Fig.~\ref{fig:gid_scheme}.
    This enables to detect diffraction from a structure that is only a single molecule thick and weakly reflective, 
    such as a Langmuir monolayer.
    The earliest interpretations of diffraction patterns from linear amphiphilic molecules at liquid surfaces are comprehensively summarized in~\cite{kjaer1994some}.
    Langmuir monolayers are two-dimensional polycrystalline systems, 
    and their diffraction patterns consist of elongated vertical diffraction peaks, known as Bragg rods. 
    The shape of the Bragg rods in the lateral direction reflects the finite size and the shape of crystallites, 
    provided the instrument resolution is sufficient, 
    while their shape in the vertical direction
    encodes the structure of amphiphilic molecules constituting the monolayer.
    Traditionally, the analysis of diffraction patterns involves determination of the positions and widths of Bragg rods to infer molecular ordering.
    However, in cases where the molecular organization becomes more complex, such simplified approaches are insufficient, and physics-based simulations of the full two-dimensional diffraction pattern are required.
   
    The present study is focused on the theoretical framework for calculating full two-dimensional GID patterns based on the distorted wave Born approximation (DWBA).
    Although similar theoretical approaches have been reported in the literature~\cite{pignat2007grazing,chuev2020theoretical},
    they typically address only specific monolayer configurations and neglect dynamic effects of X-ray scattering,
    thereby overlooking phenomena such as total external reflection. 
    Furthermore, 
    these methods often lack generality in accounting for statistical deviations of molecules within the crystal lattice.
    In some cases, they are inadequate for comprehensive diffraction analysis --- an issue that became particularly evident in our recent study~\cite{nikolaev2025probing}.
    
    To ensure flexibility in describing the monolayer structure,
    we derive it anew from basic principles.
    Then we consider three different states of the Langmuir monolayer.
    The first case is a typical solid-state monolayer.
    In this case, the diffraction pattern is calculated by treating the monolayer as a two-dimensional "powder" crystal.
    The second case is a Langmuir monolayer with mosaicity distributed around the vertical direction. 
    In our previous study, 
    we demonstrated how Langmuir monolayers can undergo this type of structural disordering under specific thermodynamic  conditions~\cite{nikolaev2025probing}.
    In the present work, 
    we further analyze the diffraction characteristics of the collapsed monolayer state.
    The diffraction pattern for this state is characterized by a horizontally broadened peak, 
    indicative of the loss of long-range positional order.
    That is, the monolayer is no longer crystalline, 
    and the observed diffraction occurs from short-range correlations between neighboring molecules, indicating a paracrystalline state.
    To model this behavior, 
    we incorporated a short-range order (SRO) correlation function into our theoretical framework.
    The results show that the proposed approach accurately reproduces diffraction patterns across a broad spectrum of lateral disorder, 
    ranging from well-ordered crystalline phases to structurally disordered collapsed states.
    This study provides a comprehensive theoretical derivation,
    numerical simulations, 
    and a quantitative comparison with experimental data,
    thereby validating the applicability of the model to complex soft matter 2D systems.

\section{Theory}

    The scattering of X-rays in Langmuir monolayers can be modeled using the Helmholtz equation and the scalar approximation:
    \begin{equation}
        (\Delta + k_0^2)E(\bm r)=V(\bm r)E(\bm r)
        .
        \label{eq:wave}
    \end{equation}
    Here, $k_0$ represents the wave number in vacuum, $E(\bm r)$ denotes the electric field in the scalar approximation, 
    and $V(\bm r)$ represents the potential of the scattering structure. 
    In the context of electromagnetic radiation scattering, 
    this potential assumes the form $V = -k_0^2 \chi (\bm r)$, 
    where $\chi(\bm r)$ is the dielectric susceptibility of the media.
    It should be noted that the Helmholtz equation is written in the scalar approximation. 
    This approximation effectively describes of X-ray scattering, 
    which can be justified as follows. 
    In the X-ray range, the real part of the refractive index, $n$, 
    for most chemical elements deviates from unity only in the fifth decimal place~\cite{als2011elements}. 
    Consequently, polarization-related effects occur far from grazing-incidence angles, 
    since the Brewster's angle $\arctan (n_1/n_2) \sim 45^{\circ}$, 
    where $n_1$ and $n_2$ are the refractive indices of two bordering media.
  
    In the case of planar continuous medium, 
    where $V \equiv V(z)$, the Helmholtz equation has a simple exact solution~\cite{daillant2008x}:
    \begin{equation}
        E(\bm r)=
        \left[
            R(z) e^{ik_zz}+T(z) e^{-ik_zz}
        \right]
        e^{-i\bm{k}_\parallel \cdot \bm{r}}
        ,
        \label{eq:planar_solution}
    \end{equation}
    where $\bm{k}_\parallel$ is the projection of the wave vector onto the horizontal plane, 
    and $k_z$ is the vertical component of the wave vector. 
    The complex amplitudes, representing the reflection and transmission coefficients, $R(z)$ and $T(z)$, respectively,
    can be obtained from the continuity conditions of the electric and magnetic fields at the interface~\cite{daillant2008x}. 
    Due to the symmetry of Eq.~\eqref{eq:wave} for planar medium, 
    the projection of the wave vector remains invariant~\cite{landau2013electrodynamics}: 
    $\bm{k}_\parallel = \text{const}$, 
    while the vertical component, $k_z$, 
    follows the spherical dispersion relation:
    $k^2_z = (1+\chi_0) k_0^2 - k^2_\parallel$.

    Equation~\eqref{eq:planar_solution} describes specular reflection, 
    wave propagation in planar medium, 
    and the amplitude of the evanescent wave. 
    The latter is crucial in the context of diffraction under grazing-incidence geometry, 
    as measurements are performed near the total external reflection condition. 
    However, Eq.~\eqref{eq:planar_solution} is evidently insufficient for describing diffraction in a Langmuir monolayer, 
    since the planar medium approximation fails to account for the lateral structure of the ensemble of amphiphilic molecules.
    It should be noted that the surface is imperfect and exhibits some three-dimensional scattering structure, 
    i.e $V(\bm r) \not\equiv V(z)$.
    In this case, the Distorted-wave Born approximation (DWBA) is applied~\cite{sinha1988x,kaganer1995bragg}.
    
    The DWBA can be formulated both in terms of quantum mechanics~\cite{sinha1988x} and within the framework of electrodynamics of the continuous media~\cite{kaganer1995bragg}. 
    In~\cite{dmitrienko_kaganer_1990}, an elegant result was demonstrated: 
    DWBA can be derived from the reciprocity theorem. 
    By applying the reciprocity theorem and considering the scalar approximation, 
    the scattering amplitude $f$ of the wave can be determined:
    \begin{equation}
        f \propto 
        \int_{\mathbb{R}^3}
        E_f V_\delta E_i \, d^3r
        .
        \label{eq:f}
    \end{equation}
    This expression is referred to in the literature as DWBA.
    Using the solution~\eqref{eq:planar_solution} of the Eq.~\eqref{eq:wave} for the incident and scattered waves,
    together with the scattering potential included in the Eq.~\eqref{eq:wave}, 
    the DWBA method allows the potential to be divided into two components:
    the potential of an ideal planar system and the potential of a non-ideal system, 
    treated as a perturbation, 
    i.e., the resulting potential is $V(\bm r) = V_\delta(\bm r) + V(z)$. 
    Thus, in our system, the monolayer itself is regarded as a perturbation $V_\delta(\bm r)$. 

    Let us consider fields $E_i$ and $E_f$ defined by solution~\eqref{eq:planar_solution}. 
    Substituting these into~\eqref{eq:f}, we obtain
    \begin{multline}
        f\propto 
        R_i R_f \hat{V_\delta}(\bm{Q}_\parallel,
            {k}_{z,i}+{k}_{z,f}) 
        +
        R_i T_f \hat{V_\delta}(\bm{Q}_\parallel,
            {k}_{z,i}-{k}_{z,f})
        +\\+
        T_i R_f \hat{V_\delta}(\bm{Q}_\parallel,
            -{k}_{z,i}+{k}_{z,f}) +
        T_i T_f \hat{V_\delta}(\bm{Q}_\parallel,
            -{k}_{z,i}-{k}_{z,f})  
        .
        \label{eq:f_full}
    \end{multline}
    Here, $\bm{Q_\parallel} = -\bm{k}_{\parallel,i}+\bm{k}_{\parallel,f}$ is the projection of the reciprocal space vector onto the surface. 
    Thus, the scattering amplitude is a sum over four scattering channels, 
    where the Fourier transforms of the perturbation $V_\delta$, 
    calculated at the corresponding points in reciprocal space, 
    act as weighting factors.

    The diffraction intensity distribution is calculated using the well-known formula for the differential scattering cross-section
    \begin{equation}
        \dfrac{d\sigma}{d\Omega} 
        \propto
        ff^*
        . 
        \label{eq:ff}
    \end{equation}
    Thus, the expression for the differential scattering cross-section explicitly consists of sixteen terms, 
    each of which comprising a set of amplitude coefficients, $R$ and $T$, 
    and a single factor $\hat{V_\delta} \hat{V_\delta^*}$. 
    The dynamical effects of evanescent wave scattering are accounted for by the amplitude coefficients, 
    while the factor $\hat{V_\delta} \hat{V_\delta^*}$ is directly responsible for diffraction.
    
    Let us focus on the amplitude coefficients. 
    The reflection coefficient in the X-ray range is significantly lower than in the optical range. 
    This can be confirmed using the Fresnel equations. 
    The reflection coefficient asymptotically decreases as the square of the refractive index decrement, 
    i.e., $|R|^2 \sim \delta$, where even for heavy chemical elements within the X-ray range, $\delta \sim 10^{-5}$. 
    Therefore, in calculations, the fifteen terms containing amplitudes $R$ out of the total sixteen terms arising during the derivation are often neglected, 
    resulting in the simplified expression
    \begin{equation}
        \dfrac{d\sigma}{d\Omega}
        \propto
        |T_f|^2 \hat{V_\delta}(\bm Q) \hat{V}_\delta^*(\bm Q)
        .
        \label{eq:main}
    \end{equation}
    This is even more applicable to organic systems, 
    as the decrement of the refractive index for carbon is approximately $\delta \sim 10^{-6}$ in the X-ray energy range. 
    The coefficient $T_i$ can be disregarded since it remains constant at a fixed incident angle, $\alpha_i$, 
    and does not influence the appearance of the diffraction pattern beyond a constant factor. 

    Consider a case where the positions and orientations of the molecules in the monolayer are random. 
    In this case, the scattering potential, $V_\delta$ represents a realization of a certain stochastic process $\{ V_\delta \}$. 
    If we assume that scattering potential $\{V_\delta\}$ is ergodic, 
    meaning that averaging over realizations is equivalent to spatial averaging, 
    we can invoke the Wiener-Khinchin theorem~\cite{wiener1930generalized,khintchine1934korrelationstheorie}.
    This theorem states that the modulus squared of a realization of a stochastic process is equal to the Fourier transform of the autocorrelation function of the process itself:
    \begin{equation}
        \dfrac{d\sigma}{d\Omega}
        \propto
        |T_f|^2 \mathcal{F}\{ C(\bm{r}) \},
        \quad
        C(\bm{r}) = \langle V_\delta(\bm{r}'-\bm{r}) V_\delta^*(\bm{r}') \rangle
        .
        \label{eq:stochastic}
    \end{equation}
    The angle brackets indicate averaging.
    Remarkably, Eq.~\eqref{eq:main} does not involve averaging.
    The averaging in Eq.~\eqref{eq:stochastic} arises from the assumption that the perturbation is ergodic, i.e., 
    averaging over one complete realization of the perturbation is equivalent to averaging over many realizations.
    The factor $\langle \hat{V_\delta} \hat{V_\delta^*} \rangle$ allows for flexibility in the choice of the autocorrelation functions, 
    that different types of structural arrangements can be considered.
    
\begin{figure}[t!]
        \centering
        \includegraphics[]{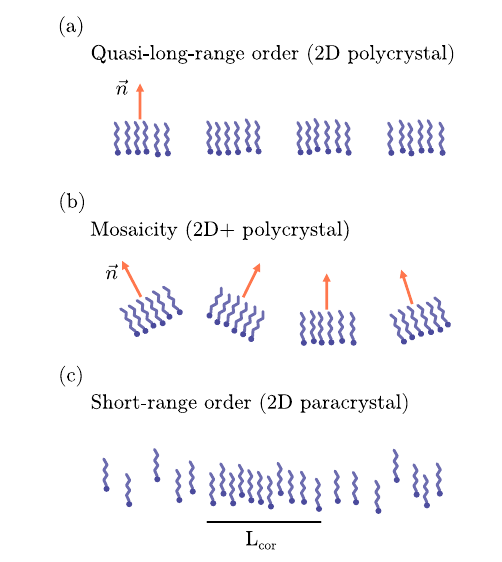}
        \caption{
            Schematic representation of lateral disordering in Langmuir monolayers, typical for the following models:
            (a) quasi-long-range order model in 2D polycrystal, 
            (b) the model of mosaicity around the vertical direction with tilted crystallites, $\vec{n}$ represents normal vector, 
            (c) short-range order model in 2D paracrystal with correlation length $L_\text{cor}$.
        }
        \label{fig:disorder_models}
    \end{figure}

    With this in mind, 
    let us consider different types of lateral ordering of the monolayer: 
    quasi-long-range order model in 2D polycrystal, 
    the model of mosaicity around the vertical direction with tilted crystallites 
    and short-range order model in 2D paracrystal, 
    which are schematically shown in Fig.~\ref{fig:disorder_models}. 

\subsection{Quasi-long-range order model} 
    
    When considering various types of ordering, we begin with a simple case of solid-state monolayer, 
    specifically, 
    monolayer with quasi-long-range order (QLRO). 
    In this case, 
    the monolayer is considered within the approximation of a 2D polycrystal, 
    as schematically illustrated in Fig.~\ref{fig:disorder_models}(a).
    In a monolayer with QLRO, 
    the positions of the molecules, $\bm{R}_m$, 
    are determined by the vectors of elementary translations and deviations of the molecules from their ideal positions, 
    which follow a three-dimensional independent normal distribution:
    \begin{equation}
        \bm{R}_m 
        = 
        \bm{r}_m 
        + 
        \bm{\delta}_{m}
        ,
        \label{eq:lattice}
    \end{equation}
    where $\bm{r}_m = m_1\bm{a}_1 + m_2 \bm{a}_2$, 
    ${\bm{a}}_{1,2}$ 
    are the basis vectors,
    $\mbox{\bm{$\delta$}}_{m} 
    \sim 
    \mathcal{N}(0,\sigma_x)
    \times
    \mathcal{N}(0,\sigma_y)
    \times
    \mathcal{N}(0,\sigma_z)$ 
    represent the molecular deviations, 
    where $\sigma_{x,y,z}$ 
    are the standard deviations along corresponding directions.
    Let us assume for definiteness, 
    that the number of lateral translations along each basis vector in a single crystallite is
    $M = M_1M_2$. 
    Introducing statistical deviations into the system necessitates considering stationary ergodic processes. 
    This implies that for  a sequence of independent identically distributed random variables, 
    ensemble averaging over realizations of the system is equivalent to a summation over positions. 
    
    Now, let us focus on the perturbation $V_\delta $. 
    To begin, we use the kinematic theory of diffraction to describe coherent scattering from a single crystallite. 
    The crystal lattice is defined by the vector $\bm{R}_m$ in real space, 
    while $\bm{Q}$ represents the reciprocal lattice vector. 
    The Fourier transform of the perturbation $V_\delta $ 
    is expressed in the form:
    \begin{equation}
        \hat{V}_\delta 
        = 
        \sum_{m} 
        \hat{F}_m(\bm{Q}) 
        e^{
        -i \bm Q\cdot\bm{R}_m
        }
        ,
        \label{eq:scat_potential}
    \end{equation}
    where $\hat{F}_m(\bm{Q})$ is the form factor of the $m$-th molecule.
    In Eq.~\eqref{eq:scat_potential}, 
    the summation runs over all molecules in a single crystallite.

    Consider some model details specific to the Langmuir monolayer.
    Assume that all molecules in the crystal are identical and oriented uniformly in space. 
    In this case, 
    the form factors of all molecules are equal:
    $\hat{F}_m(\bm Q) = \hat{F}(\bm Q)$. 
    The form factor of a molecule within a Langmuir film is represented by a disk in reciprocal space~\cite{kaganer1999structure} 
    with a thickness of $2\pi/H$, 
    where $H$ is the characteristic height of the molecule along its main axis. 
    This simplified representation arises from the assumption 
    that the electron density of the molecule 
    is uniformly distributed within a cylinder of height $H$. 
    In reciprocal space, 
    the molecules contribute to the intensity only near the plane perpendicular to the main axis of the molecule. 
    Additionally, 
    the characteristic features of X-ray scattering from a single unit cell of the crystallite should be considered, 
    which is described by the factor $\hat{S}(\bm Q)$. 
    It is convenient to describe the crystal structure of the Langmuir monolayer by a rectangular centered unit cell:
    a cell with one molecule at the origin and another at the center.
    In general, 
    the lateral translation of this unit cell corresponds to an oblique hexagonal arrangement of the molecules on the surface.
    In this case, 
    the structure factor is 
    \begin{equation}
        \hat{S}(\bm{Q}) 
        = 
        1
        +
        \exp{[
        -i \bm{Q}
        \cdot
        ({\bm{a}}_1
        +
        {\bm{a}}_2)
        /2
        ]}
        .
        \label{eq:sf}
    \end{equation}
    This defines the local crystal structure of the Langmuir monolayer. 

    Thus, the general expression (\ref{eq:stochastic}) in the QLRO model can be explicitly expressed in terms of the structure and form factors as follows: 
    \begin{equation}
        \dfrac{d\sigma}{d\Omega}
        \propto
        \left\langle
        |T_f(Q_z)\hat{S}(\bm Q_\parallel)\hat{F}(\bm{Q})|^2G(\bm Q)
        \right\rangle_\text{cyl}
        \label{eq:LRO}
    \end{equation}
    Here, $\langle\cdot\rangle_\text{cyl}$ denotes averaging in cylindrical coordinates.
    This accounts for the incoherent contribution to the diffraction pattern from randomly oriented crystallites uniformly distributed on a surface.
    Function $G$ is a correlation term,
    and according to a Wiener-Khinchin theorem, it involves averaging over statistical realizations:
    \begin{equation}
        G(\bm Q) = 
        \sum\limits_{ mn } 
        \left\langle
        e^{-i \bm{Q}\cdot\bm{R}_m} 
        e^{i \bm{Q}^*\cdot\bm{R}_n}
        \right\rangle
        .
        \label{eq:correlator}
    \end{equation}
    The indices $m, n$ range over the nodes of a single crystallite lattice.
    Given the normal distribution and the statistical independence of the deviation vectors $\bm{\delta}_m$ in Eq.~\eqref{eq:lattice}, 
    a well-known result follows~\cite{debye1915scattering,farrow2009relationship}:
    $
        G(\bm Q) = J(\bm{Q}_\parallel) \eta(\bm{Q})
    $;
    where $J(\bm{Q}_\parallel)$ is a two-dimensional interference function:
    \begin{equation}
        J(\bm{Q}_\parallel)
        =
        \frac{\sin^2{M_1\kappa_1}}{\sin^2{\kappa_1}} 
        \frac{\sin^2{M_2\kappa_2}}{\sin^2{\kappa_2}}
        ,   
        \label{eq:debye}
    \end{equation}
    with $ \kappa_{1,2} = \bm{Q}_\parallel \cdot 
    \bm{a}_{1,2}/2$.   
    In the context of classical powder diffraction, 
    the explicit form of the interference function, 
    which is given by the Debye formula~\cite{debye1915scattering,farrow2009relationship}, 
    is of particular significance.  
    The function $J(\bm{Q}_\parallel)$ describes the interference of scattered waves on the crystal lattice and determines the position of Bragg rods in the diffraction map. 
    The scattered radiation aligns along specific lines in reciprocal space that are perpendicular to the surface, 
    referred to Bragg rods. 
    The last term $\eta(\bm{Q})$ in the correlation term $G(\bm Q)$ is simply a Debye-Waller factor accounting for the position fluctuation of the molecules:
    \begin{equation}
        \eta(\bm{Q}) 
        = 
        \prod_{i=\{x,y,z\}}
        \exp{(-Q_i^2\sigma_i^2/2)}
        .      
    \end{equation}
    It handles the reduction in diffraction intensity due to fluctuations of $\bm{\delta}_m$ molecular positions in the lattice. 
    Despite the fact that the monolayer is two-dimensional,
    there are vertical components of deviation in the system.

    Thus, Eq.~\eqref{eq:LRO} describes the distribution of diffraction intensity observed from a polycrystalline monolayer.
    All the standard considerations regarding X-ray scattering on a monolayer, 
    as outlined in classical works, remain valid, 
    with the only adjustment being the amplitude $T_f$, 
    which incorporates the dynamic effects of evanescent wave scattering: specifically, 
    the Yoneda peak  near the critical angle~\cite{yoneda1963anomalous,vineyard1982grazing}. 
    It is important to emphasize that the case of statistical deviations of molecules from ideal positions in the crystal lattice 
    is modeled as a three-dimensional independent normal distribution, 
    that describe molecular disorder both within the plane of the structure and vertically. 
    The proposed model is universal and flexible. 
    In the frame of this model, it is possible to define the statistics of molecular deviations using a different   distribution function. 
    Consequently, 
    the entire theory described above applies to arbitrary cases. 

\subsection{Model of mosaicity around the vertical direction}

    A typical GID pattern from Langmuir monolayer characterises a set of Bragg rods oriented perpendicularly to the monolayer surface. 
    This corresponds to the scattering on a perfectly two-dimensional polycrystalline structure.
    However, under certain conditions~\cite{daillant1999short},
    the overcompressed monolayers buckle into the third dimension.
    Buckling has been observed in polymerized monolayers 
    (see references~\cite{bourdieu1994buckling,fontaine1997spontaneous,gourier2001structure}).
    The buckling mechanisms proposed in these works differ markedly from one another.
    However, they all involve stiffening the monolayer through various molecular processes.
    This stiffening prevents the overcompressed monolayer from collapsing. 
    Instead, the monolayer releases the mechanical stress through its corrugation.
    In our recent studies~\cite{nikolaev2025probing},
    we also observed data that can be explained by a loss of the horizontal order.
    We will discuss this data later. 
    For now, 
    let us focus on how a partial loss of horizontal positional order affects the GID pattern.
    
    We consider a corrugated monolayer by introducing the mosaicity around the vertical direction.
    Namely, we consider the flat crystallites in a two-dimensional monolayer to be slightly deviated from their ideal horizontal position
    [see Fig.~\ref{fig:disorder_models}(b)]. 
    If the system comprises a sufficiently large number of randomly distributed 2D crystallites, 
    the central limit theorem asserts that the overall distribution will approximate a normal distribution. 
    The inclination of each crystallite with respect to the horizon is represented by the angle $\beta$, 
    which is defined as the angle between the normal to the subphase surface and the normal to the crystallite.
    The inclination $\beta$ is then follows the normal distribution with the probability density:
    $\rho(\beta) \sim
    \exp[
    -\beta^2/2\sigma_\beta^2
    ]$, 
    where $\sigma_\beta$ is the standard deviation of the inclination.
    The GID pattern is formed by averaging the pattern calculated using QLRO [cf. Eq.~\eqref{eq:LRO}] across all crystallite orientations:
    \begin{equation}
        \dfrac{d\sigma}{d\Omega} 
        \propto
        \int_{0}^{\pi} 
        d\beta 
        \rho(\beta) 
        \sin{\beta}
        \int_{0}^{2\pi} 
        d\theta 
        |f(\mathbf{Q})|^2
        .
        \label{eq:vertical_mosaicity}
    \end{equation}
    Expression~\eqref{eq:vertical_mosaicity} describes the GID pattern of a Langmuir monolayer, 
    taking into account the disruption of the horizontal order within a 2D polycrystal.
    Note that such a simplistic model disregards position-orientation correlations by averaging as $\braket{|f|^2}$
    instead of averaging as
    $|\braket{f}|^2$.
    In other words, this model does not distinguish between a continuously corrugated layer and a mosaic crystal with abrupt orientation changes between neighboring crystallites.
    Therefore, there is no interference between the waves scattered by different crystallites. 
    As a result, only the local changes in intensity distribution around a Bragg rod are considered in (\ref{eq:vertical_mosaicity}).
    
\subsection{Short-range order model}
    \label{sec:SRO}
    
    Previously,
    we studied the disordering in the specific collapse mode,
    characterized by a corrugation of monolayer,
    using
    the model of mosaicity around the vertical direction.  
    Collapse is well-known to induce the formation of three-dimensional aggregates. 
    But even under conditions of global disordering, the persistence of local correlations at the surface could be expected. 
    This is manifested in GID patterns
    as a broadening of the peaks due to a progressive loss of correlation between particle positions, e.g., as in~\cite{dittrich2011physical}.
    
    Quantitative description of local positional correlations or short-range order (SRO) within Langmuir monolayers remains a challenging task.
    Two principal approaches have been developed here to resolve this problem.
    The first derives from the statistical theory of simple liquids and is based on the hard-sphere model,
    which provides a well-established framework for analyzing fluid structure~\cite{croxton1974liquid,balescu1975equilibrium}.
    Within this model, there are reliable approximations for the pair density correlation function, $g(r)$,
    which enable the structure factor for a liquid to be calculated.
    For such a model a good approximation for the direct correlation function,
    related to the $g(r)$ function by the integral Ornshtein-Zernike expression,
    is the Percus-Yevick approximation.
    In this formulation,
    the structure factor depends only on the hard-sphere diameter $D$
    and the volume fraction of particles~\cite{croxton1974liquid, balescu1975equilibrium}.
    
    The second approach is based on the paracrystal model, which was extensively developed and popularized during the 1950s and 1960s by Hosemann~et.al.~\cite{hosemann1962direct}.
    In this framework, the loss of long range positional order within one-dimensional lattice of nodes is introduced in a cumulative way from point to point through the knowledge of the distance probability between two neighboring nodes
    (see, for example~\cite{vainshtein1966diffraction, renaud2009probing}, and references therein).
    Given that Langmuir monolayers consist of randomly oriented microcrystallites,
    the paracrystal model offers a natural and effective means for describing their partial positional disordering.
    In our work, we adopt modern extensions of the paracrystal model which incorporate statistical descriptions of lattice distortions and allow for more accurate simulations of GID patterns.
    
    To describe SRO in a Langmuir monolayer,
    we start from one-dimensional model and then extend it to the two-dimensional case, 
    following approaches similar to those used in~\cite{buljan2012grazing, nikolaev2019grazing}.
    Consider an ideal linear lattice with the parameter $\bm{a}$.
    Although it is a one-dimensional model,
    consider that the deviation vectors $\bm\delta_i$ exist in three dimensions.
    In other words, molecules can deviate from an ideal linear arrangement.
    As in Eq.~\ref{eq:lattice},
    the deviations are stochastic and normally distributed. 
    The representation of the molecular positions within the one-dimensional SRO model is that the deviation of the $m$-th cell is the sum of all previous realizations of $\bm{\delta}_{i<m}$.
    The equation for the real positions, $\bm{R}_m$, can be written as a cumulative sum:
    \begin{equation}
        \bm{R}_m 
        =
        \bm{r}_m
        +
        \sum_{i=0}^m
        \bm{\delta}_{i}
        .
        \label{eq:lattice_cumulative}
    \end{equation}
    Note that the average position remains at the ideal position due to statistical independence of $\bm{\delta}_i$, i.e.,
    $\braket{\bm{R}_m} = m\bm{a}$.
    However, 
    the correlation term $G(\bm Q)$
    involves an average $\braket{\bm{R}_m\bm{R}_n}$ 
    [cf. Eq.~\eqref{eq:correlator}].
    which grows with respect to $|m-n|$.
    As schematically shown in Fig.~\ref{fig:disorder_models}(c), 
    such systems are characterized by a correlation length $L_\text{cor}$ 
    that sets the scale over which positional correlations between molecules persist 
    and beyond which the order gradually decayes due to the accumulation of deviations.

    By substitution of Eq.~\eqref{eq:lattice_cumulative} in Eq.~\eqref{eq:correlator} and introducing the phase term for a lattice translation
    $\xi(\bm{Q}) = \exp(-i\bm{Q}\bm{a})$,
    one can calculate an explicit form of the correlation term $G$ for 1D SRO model~\cite{buljan2012grazing}:
    \begin{equation}
        G(\bm{Q}) =
        2\text{Re}
        \left\{
            \frac{M}{1-\xi\eta} - 
            \xi\eta\frac{1-(\xi\eta)^M}{(1-\xi\eta)^2}
        \right\}
        -M
        .
        \label{eq:SRO_1D}
    \end{equation}
    Here, the Debye-Waller factor is calculated for the projection of vector $\bm Q$ onto $\bm a$:
    $
        \eta(\bm{Q}) = 
        \exp(-Q_a^2\sigma^2/2)
    $.

    \begin{figure}[t!]
        \centering
        \includegraphics{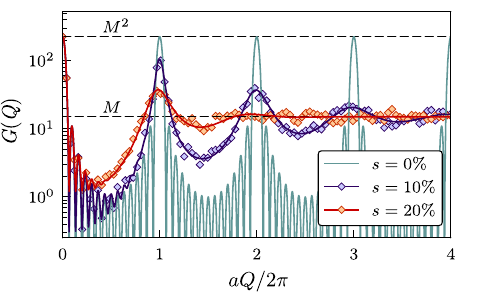}
        \caption{
            Correlation term $G(Q)$ computed for different values of
            unitless deviation parameter $s = \sigma/a$.
            The deterministic calculations from Eq.~\eqref{eq:SRO_1D} are shown with solid lines. 
            Monte Carlo simulations of Eq.~\eqref{eq:correlator} are shown with markers for comparison.
            }
        \label{fig:G}
    \end{figure}

    Thus, Eq.~\eqref{eq:SRO_1D} describes the interference of waves scattered on a one-dimensional chain with accumulated disorder.
    Its behavior does not read as straightforwardly as a Laue function in Eq.~\eqref{eq:debye}.
    Let us point out its features.
    First, it can be shown~\cite{nikolaev2019grazing} that
    $G(0) \to M^2$ and for a finite $\sigma$: $G(\infty) \to M$.
    Naturally, for $\sigma \to 0$,
    the correlation term $G$ should behave as the interference function: $G(\bm Q)\to J(\bm Q)$.
    To demonstrate this,
    we calculated $G(Q)$ for different distributions of $\bm{\delta}_i$
    (see Fig.~\ref{fig:G}).
    For convenience,
    we introduce the unitless parameter $s = \sigma/a$,
    it represents the standard deviation in terms of fractions of the lattice parameter.
    The spatial frequency is also given as unitless parameter $aQ/2\pi$. 
    Fig.~\ref{fig:G} shows three correlation curves:
    one for a case of an absence of deviations ($s = 0$)
    and two others for deviations constituting 10\% and 20\%.
    All curves are calculated for $M = 15$.
    The curves calculated for finite values of $s$ are compared with the results from the simple Monte Carlo simulations, 
    represented as markers.
    These simulations are performed by directly computing
    Eq.~\eqref{eq:correlator} for a cumulative sum of realizations of $\bm{\delta}_i$ and averaging over a $100$ of such computations.
    The deterministic calculations of $G$ are consistent with Monte Carlo simulations.
    
    Indeed, the curve for $s = 0$ behaves like an interference function: 
    it is equal to $M^2$ at each node and the width of the peaks is determined by the number $M$.
    As the parameter $s$ increases,
    the intensity of the higher-order peaks diminishes, while their width increases.
    This broadening of the peaks represents a qualitative difference with QLRO,
    as it indicates the finite correlation length characteristic of a paracrystal.
    At $s =20\%$,
    only the first-order peak remains visible.
    The physical interpretation is that only neighboring molecules with $|\bm{R}_m-\bm{R}_n| < L_\text{cor}$ contribute coherently to scattering, resulting in interference,
    while scattering from more distant molecular pairs  contributes to a background level proportional to the total number of scatterers $M$.
    Thus, the parameter $s$ serves as a parameter of continuous transition from a crystalline to a paracrystalline state.

    As discussed in~\cite{buljan2012grazing}, 
    the two-dimensional correlation function can be approximated as a product of two one-dimensional correlation functions,
    for translations along $\bm{a}_1$ and $\bm{a_2}$.
    Taking this into account, and incorporating vertical fluctuations through a standard Debye-Waller factor along $Q_z$,
    one can infer the correlation function for the SRO model:
    \begin{equation}
        G(\bm{Q}) = G_1(\bm{Q}_\parallel)G_2(\bm{Q}_\parallel)\eta(Q_z).
        \label{eq:SRO_2D}
    \end{equation}
    Such partitioning ensures consistent modeling of out-of-plane disorder 
    while preserving the description of lateral correlations through two-dimensional correlation function.
    Finally, the calculation of the scattering intensity can be done similarly to QLRO, by taking Eq.~\eqref{eq:SRO_2D} into Eq.~\eqref{eq:LRO}.
    For a polydisperse system, 
    Eqs.~\eqref{eq:LRO} and \eqref{eq:SRO_2D} acquire a more complex form, 
    in which each term in the structure factor is multiplied by its individual form factor, 
    specific to the molecular species involved.
    However, for all subsequent simulations, we assume a monodisperse system.
    
\section{Experimental details}
    
    The next step involves validating the three theoretical models using experimental data. 
    These three structural states of the monolayer --- polycrystalline, corrugated, and collapsed --- represent distinct stages of lateral disordering and
    were all observed in the previous study~\cite{nikolaev2025probing}. 
    Utilizing these data is particularly advantageous, 
    as they were obtained under similar experimental conditions using monolayers of identical  chemical composition. 
    A detailed description of the experimental procedures and corresponding interpretations are provided  in~\cite{nikolaev2025probing}.
    For the reader's convenience, 
    a brief summary of the experimental setup and sample parameters is included below.

    The GID measurements were performed at the bending magnet beamline "Langmuir"~\cite{yakunin22spectral} of the Kurchatov synchrotron radiation source.
    The beamline is equipped with a Langmuir trough.
    In all the diffraction measurements, the incidence angle was fixed at
    $\alpha_i = 0.8\alpha_c$,
    where $\alpha_c$ is the critical angle of total external reflection for water.
    The incident synchrotron beam was monochromated to photon energy of 13~keV. 
    GID data were collected by scanning with a strip Mythen2 X 1k position-sensitive 1D detector equipped 
    with the Soller slits to compensate for the parallax on a large illuminated area.

    To form a Langmuir monolayer,
    we spread a solution of the arachidic acid  in chloroform 
    (concentration of 0.59~mg/ml) on the surface of Ce(NO$_3$)$_3\, \cdot \,$6H$_2$O salt solution 
    (concentration of 0.08~g/l). 
    The solvent was allowed to evaporate for 15 min.
    After that the layer was compressed to a working surface pressure. 
    The effect of Ce$^{3+}$ ions on the structural ordering of the arachidic acid monolayer was examined at two temperatures (21 and 23 $^\circ$C) and at different surface pressures.
    We also applied X-ray standing wave (XSW) technique to monitor the location of cerium atoms during the complicated process of molecular reorganization of arachidic acid monolayer. In addition Brewster angle microscopy (BAM) measurements were performed to visualize the changes in monolayer morphology at different stages of compression.

    We choose for QLRO model simulations experimental GID data, obtained for arachidic acid monolayer compressed to a surface pressure of 20~mN/m at a temperature of 23$^\circ$C, 
    that corresponds to solid phase monolayer (the first sample). 
    The second sample is a monolayer compressed to a surface pressure beyond the collapse point $\pi = 52$~mN/m at a temperature of $T = 21^\circ$C. 
    An interesting case of structural arrangement has been observed under these conditions. 
    Despite being overcompressed,
    the monolayer maintained high degree of structural ordering. 
    The mechanical stress seems to release through the corrugation of the monolayer. 
    These GID data were selected to test the mosaicity model. 
    The arachidic acid monolayer at $\pi = 52$~mN/m and $T = 23^\circ$C exhibit typical behavior, 
    observed in collapsed films. 
    Although the monolayer loses its structural integrity, 
    a faint diffraction peak remains clearly visible in experimental GID pattern. 
    As an explanation of this observation we hypothesize the formation of a paracrystalline structures with SRO in collapsed monolayer.
    We used these GID data to test the SRO model (the third sample). 
    
\section{Experimental data, numerical simulations and discussion}
    
\begin{figure}[b!]
        \centering
        \includegraphics[]{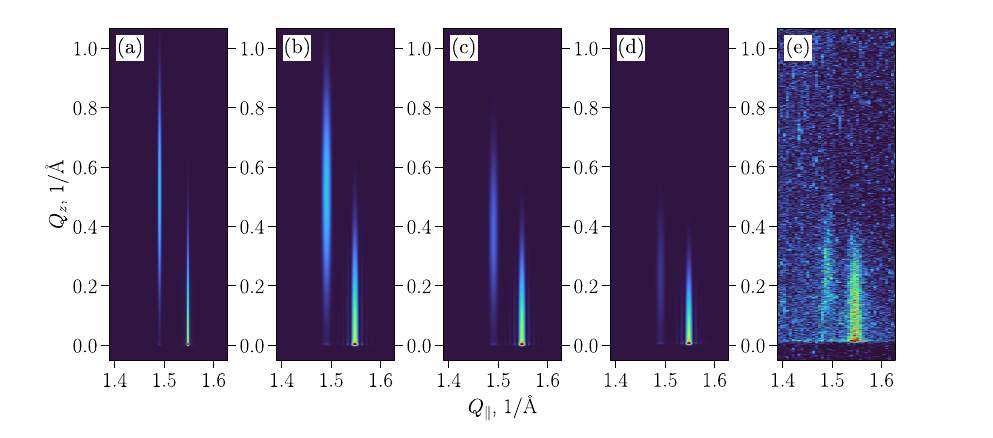}
        \caption{
            The GID pattern of a Langmuir monolayer in the polycrystalline solid-state phase.
            (a)-(e) Numerical simulations.
            Model parameters $H = 31$~\AA{},
            $a_1 = 4.93$~\AA{}, $a_2 = 8.12$~\AA{} and
            $\sigma_{x,y} = 1$~\AA.
            The simulations are performed 
            with respect to different roughness $\sigma_z$
            and the effective crystallite size $M$:
            (a) $\sigma_z=0$~\AA{}, $M=200\times200$;
            (b) $\sigma_z=0$~\AA{}, $M=80\times80$;
            (c) $\sigma_z=3$~\AA{}, $M=80\times80$; 
            (d) $\sigma_z=5.5$~\AA{}, $M=80\times80$.
            (e) Experimental data.
        }
        \label{fig:QLRO}
    \end{figure}

    The GID data and simulations based on the QLRO model are shown in Fig.~\ref{fig:QLRO}.
    Experimental data [see Fig.~\ref{fig:QLRO}(e)]
    show a standard diffraction pattern that corresponds to a Langmuir monolayer in the polycrystalline solid-state phase:
    a triplet of Bragg peaks.
    There is a $(02)$ peak at the horizon $Q_z = 0$~\AA{}$^{-1}$, 
    and two peaks, $(11)$ and $(1\overline1)$,
    appear above the horizon.
    These two peaks degenerate into one peak in the observed pattern 
    [Fig.~\ref{fig:QLRO}(e)].
    Note again that this index notation assumes a rectangular centered unit cell:
    [cf. Eq.~\eqref{eq:sf}].
    The reciprocal space coordinates of the peaks are:
    $(02)$ is at 
    $Q_\parallel = 1.548$~\AA{}$^{-1}$, 
    $Q_z = 0$~\AA{}$^{-1}$ and 
    ($11$), ($1\overline 1$) are at 
    $Q_\parallel = 1.491$~\AA{}$^{-1}$, 
    $Q_z = 0.3$~\AA{}$^{-1}$.
    From these coordinates one can directly infer the elementary cell parameters: 
    $a_1 = 4.93\pm0.03$~\AA{}, 
    $a_2 = 8.12\pm0.02$~\AA{} 
    with the lattice angle $\gamma = 90^\circ$.
    The confidence intervals are calculated using error estimates obtained by fitting the Gaussian curves to the experimental data.
    Additionally, 
    it can be inferred  that the monolayer is in the tilted phase, 
    wherein the aliphatic chains of the constituent molecules are oriented at an angle relative to the liquid-air interface,
    tilting toward the nearest neighbor (NN) direction rather than being aligned normal to the surface. 
    This is deduced from the fact that $(02)$ reflex is at the horizon, 
    while $(11),(1\overline1)$ degenerate reflection is above.
    Taking the vertical coordinate of $(11),(1\overline1)$ peak, $Q_z = 0.3$~\AA$^{-1}$, 
    the tilt angle is obtained, $\tau \simeq 13^\circ$.

    The GID simulations in Figs.~\ref{fig:QLRO}(a)--\ref{fig:QLRO}(d) are performed using the same lattice parameters, 
    but with different roughness $\sigma_z$ and different effective crystallite size $M$ defined as a number corresponds to the product of lateral translations along each basis vector.
    The parameters for simulation in Fig.~\ref{fig:QLRO}(d) are chosen to agree with the experiment 
    [cf. Fig.~\ref{fig:QLRO}(e)].
    The width of the Bragg peaks in the $Q_\parallel$ direction is determined by $M$.
    As a qualitative result from the kinematic theory,
    the larger the number of coherent scattering molecules, the narrower the Bragg peaks are. 
    This is also seen in the simulations 
    [Figs.~\ref{fig:QLRO}(a),~\ref{fig:QLRO}(b)].
    Following this qualitative reasoning further,
    the elongation of the peaks in the $Q_z$ direction is determined by the form factor of the molecule and, 
    consequently, 
    it is inversely proportional the molecular height $H$.
    However, 
    the simulations show that the size of the peaks in the $Q_z$ direction 
    is also affected by the  vertical roughness r.m.s., $\sigma_z$.
    Indeed, 
    the higher the $Q_z$, 
    the faster the decay in the $Q_z$ direction. 
    Furthermore, 
    peaks at different $Q_z$ coordinates are affected differently. 
    For example, 
    consider simulations in Fig.~\ref{fig:QLRO}(b)--\ref{fig:QLRO}(d),
    in which the $(11,1\overline1)$ peak appears at different $Q_z$. 
    These variations in peak position correspond to changes in the parameter $\sigma_z$.
    
    This is once again a trivial result from the kinematical theory.
    However, it limits the application of simple crystal structure analysis based solely on peak coordinates from the experimental data.
    For example, if one were to take the $Q_z$ coordinate of 
    $(11),(1\overline1)$ directly from the experimental data,
    $Q_z \simeq 0.3$~\AA{}$^{-1}$,
    the estimated tilt angle is $\tau \simeq 13^\circ$.
    However, if the position of the peak is taken from simulations with $\sigma_z = 0$~\AA,
    [peak $(11),(1\overline1)$ at
    $Q_z \simeq 0.51$~\AA{}$^{-1}$
    in Fig.~\ref{fig:QLRO}(a)]
    then $\tau \simeq 22^\circ$.
    Apart from the simulations, 
    the following qualitative argument can be presented.
    The peaks represent the intersection of the form factor and the Bragg rods. 
    For an elongated molecule, the form factor takes the shape of a disk in reciprocal space.
    Therefore, 
    all three peaks have the same vertical size in reciprocal space. 
    Since the part of reciprocal space below the horizon is unobservable in this scattering geometry, 
    the vertical size of the $(11),(1\overline1)$ peak should be twice that of the $(02)$ peak. 
    Furthermore, 
    the integrated intensity of $(11),(1\overline1)$ should be twice that of $(02)$.
    However, 
    none of this is observed in the experiment 
    [Fig.~\ref{fig:QLRO}~(e)].
    Thus, 
    the $(11),(1\overline1)$ peak decays faster than the $(02)$ peak in $Q_z$ direction.
    According to the simulations, 
    this is due to the vertical roughness of the monolayer.
    Therefore, 
    an accurate determination of the crystal structure parameters requires fitting the simulated diffraction patterns to the experimental data, 
    rather than relying solely on the analysis of peak positions. 
    This underscore the necessity for the physically-based GID simulations.
    
\begin{figure}[b!]
        \centering
        \includegraphics[]{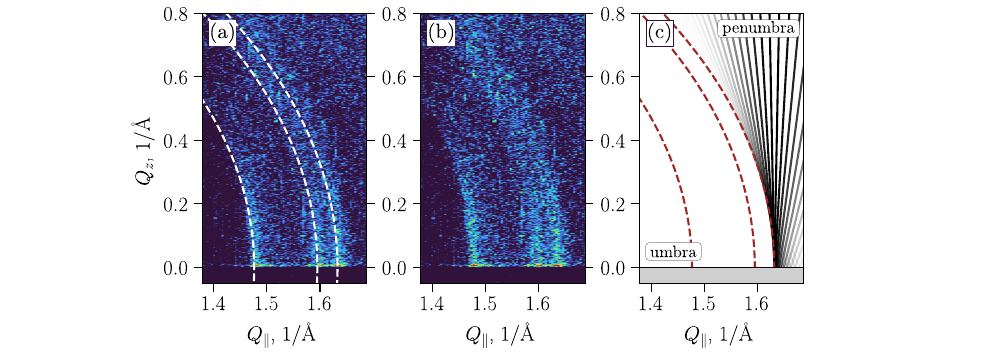}
        \caption{
            (a) Two-dimensional diffraction map from an arachidic acid monolayer formed on the surface of a cerium salt solution 
            (sample 2).
            The system was in a condensed state beyond the collapse point of the system.  
            Curved diffraction peaks are observed, and these are described by circles with a center at $Q_z = 0$~\AA{}$^{-1}$.  
            Patterns (a) and (b) are identical: 
            on the left, 
            for clarity, 
            parts of the circles describing the curvature are shown.
            (c) Schematic simulated diffraction map. 
            For clarity, 
            the simulation is shown only for the rightmost curved Bragg rod.
            The three circles on which the curved Bragg rods lie are indicated by red dashed-dotted lines. 
            The individual straight Bragg rods are shown as solid lines, 
            whose colors are represented in shades of gray according to the normal distribution.
            The rods intersect more frequently near the ring than in the region away from the ring, 
            leading to the appearance of a curved diffraction rod.
            The pattern consists of a shadow ("umbra") 
            and a partial shade ("penumbra"), 
            separated by a set of diffraction rods.
            The boundary between "umbra" and "penumbra" runs precisely along the diffraction ring 
            (red dashed-dotted line).
            }
        \label{fig:penumbra}
    \end{figure}

    The GID data obtained from the second sample are shown in Fig.~\ref{fig:penumbra}. 
    Experimental data reveal an unusual set of three curved Bragg rods 
    [see Fig.~\ref{fig:penumbra}(b)]. 
    Notably, these rods are located precisely on a circle in reciprocal space centered at the coordinate origin ($Q_z = 0$~\AA$^{-1}$) 
    [see Fig.~\ref{fig:penumbra}(a)]. 
    This is uncommon for diffraction on a 2D polycrystalline system.
    This diffraction pattern corresponds to a Langmuir monolayer in the specific collapsed state, 
    in which the overcompressed monolayer maintained its structural integrity.
    The diffraction pattern shows a peak $(02)$ at 
    $Q_\parallel = 1.476$~\AA$^{-1}$, 
    $Q_z = 0$~\AA$^{-1}$, 
    a peak ($11$) at 
    $Q_\parallel = 1.595$~\AA$^{-1}$, 
    $Q_z = 0$~\AA$^{-1}$
    and a peak ($1\overline 1$) at 
    $Q_\parallel = 1.633$~\AA$^{-1}$,
    $Q_z = 0$~\AA$^{-1}$. 
    From these coordinates one can directly infer the the elementary cell parameters: 
    $a_1 = 4.38\pm0.02$~\AA{}, 
    $a_2 = 8.52\pm0.02$~\AA{} 
    with the lattice angle $\gamma = 88.3\pm0.4^\circ$.
    As can be seen from Fig.~\ref{fig:penumbra}(b), 
    all peaks are at the horizon. 
    This observation evidences that the hydrocarbon chains of the molecules are oriented parallel to the crystallite normal, 
    i.e., the system is in the untilted phase. 
    It should be emphasized, 
    that the structure parameters determined for the second sample correspond to molecular packing, 
    which is referred in the literature as pseudo-herringbone (PHB) packing~\cite{kuzmenko1998packing}. 
    This type of molecular arrangement is characterized by dihedral angle between the planes of carbon backbone chains of approximately 40$^\circ$.
    Note, that in the case of herringbone (HB) packing this packing parameter is equal 90$^\circ$.
    For better understanding, 
    a comprehensive description of the systematic classification of molecular packing types in various fatty acid monolayers can be found in~\cite{kuzmenko1998packing}.
    Very important parameter, 
    which influences the formation of GID pattern is the inclination $\beta$, 
    the angle between the normal to a given domain surface and the normal to the subphase surface.
    This parameter characterizes the domain's orientation relative to the horizon.
    
    As a first step, we provide a qualitative description of the diffraction pattern observed in the second sample of the corrugated monolayer.
    For simplicity, 
    each domain is modeled as a flat single crystal.
    The magnitude of the inclination angle $\beta$ is assumed to follow a normal distribution,  
    while the azimuthal orientation of the domains is considered isotropic.
    Each individual domain produces a diffraction pattern consisting of straight Bragg rods inclined with respect to the horizon by an angle corresponding to the crystallite inclination.
    By averaging the diffraction contributions from all crystallites according to the normal distribution of inclination angles, 
    a composite diffraction pattern emerges, 
    as shown in Fig.~\ref{fig:penumbra}(c).
    Individual Bragg rods are shown as solid lines, 
    with their intensity represented in shades of gray according to the normal distribution.
    The rods intersect more densely near the ring than far from it, 
    producing the appearance of a curved diffraction peak.
    Notably, 
    no rods intersect the region inside the ring 
    (to the left of the arc in reciprocal space), 
    leading to a pronounced dark zone, which we named "umbra".
    Thus, diffraction pattern consists of a shadow ("umbra") 
    and a partial shade zone, which we named "penumbra". 
    This zones separate by incomplete ring of maximum intensity formed by the coherent superposition of many inclined diffraction Bragg rods.
    This feature is consistent with experimental observations 
    and provides a clear explanation for the absence of scattering signal in the shadowed region.
    
\begin{figure}[b!]
        \centering
        \includegraphics[]{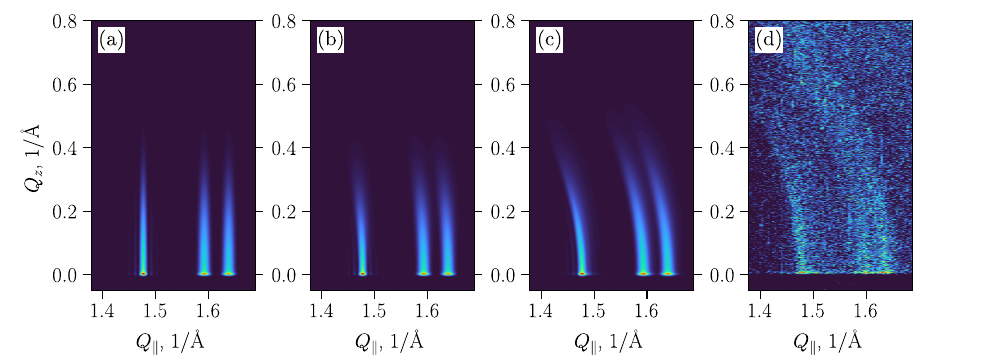}
        \caption{
            The GID pattern of a corrugated Langmuir monolayer.
            (a–c) Numerical simulations calculated for different values of the crystallite inclination variance $\sigma_\beta$:
            (a) $\sigma_\beta = 0^\circ$,
            (b) $\sigma_\beta = 3^\circ$,
            (c) $\sigma_\beta = 6.6^\circ$.
            (d) Experimental data.
            }
        \label{fig:slope}
    \end{figure}

    As the next step, 
    we provide a quantitative description of the diffraction pattern from the second sample, 
    with the results of the GID simulations based on the model of mosaicity around the vertical direction presented in Figs.~\ref{fig:slope}(a)--\ref{fig:slope}(c).
    Numerical simulations are performed using the same lattice parameters, 
    but with different values of the crystallite inclination variance $\sigma_\beta$.
    We reconstruct the parameters characterizing the lateral structural distortion in the corrugated monolayer for the second sample.
    From the best-fit results for the GID pattern 
    [Fig.~\ref{fig:slope}(d)], 
    the standard deviation of the crystallite inclination distribution was determined to be 
    $\sigma_\beta = 6.6^\circ$. 
    This implies that approximately $70\,\%$ of the domains are inclined by less
    than $6.6^\circ$ with respect to the horizon, 
    following a Gaussian distribution of orientations. 
    The best-fit simulation is shown in Fig.~\ref{fig:slope}(c).

\begin{figure}[t!]
        \centering
        \includegraphics[]{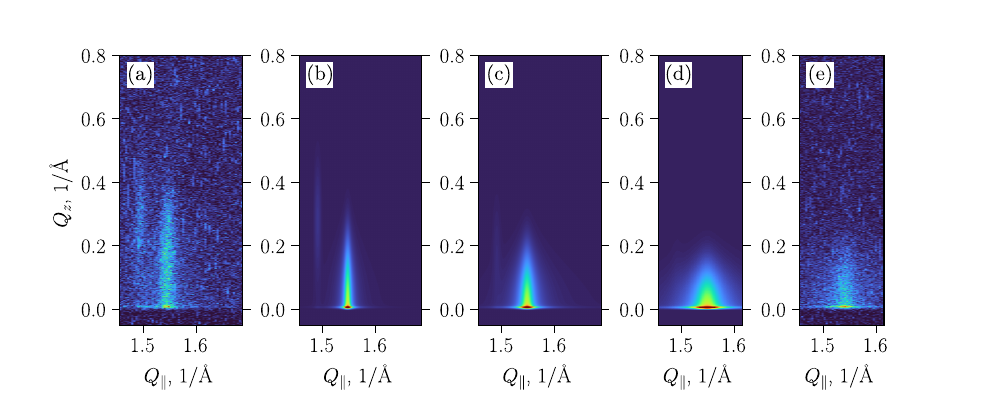}
        \caption{
            The GID pattern of a monolayer in a collapsed state.
            Experimental data (a, e) and simulations (b, c, d).
            (a) The GID pattern as measured right after the monolayer was compressed beyond the collapse point.
            (b-d) Numerical simulations based on the SRO model, which were performed for an increasing parameter $s$. 
            (e) The experimental GID pattern of a monolayer in its final collapsed state, as measured 40 hours after compression.
            The simulated patterns are arranged such that the simulation in (b) corresponds to the data in (a) and the simulation in (d) corresponds to the data in (e).
            Simulation parameters:
            $a_1 = 4.93$~\AA{},
            $a_2 = 8.12$~\AA{},
            $M = 100 \times 100$;
            (b) $\sigma_z = 6$~\AA{}, $s = 2\%$;
            (c) $\sigma_z = 8$~\AA{}, $s = 3\%$; and
            (d) $\sigma_z = 11$~\AA{}, $s = 4.5\%$.
            }
        \label{fig:SRO}
    \end{figure}

    The GID data obtained from the third sample and simulations based on the SRO model are shown in Fig.~\ref{fig:SRO}.
    Initially, the structure of the monolayer was similar to that in the first sample,
    but the structure began to degrade.
    Experimental data for the monolayer in its initial state 
    [see Fig.~\ref{fig:SRO}(a)]
    once again exhibits a standard diffraction pattern that corresponds to a Langmuir monolayer of the polycrystalline solid-state phase with molecular tilt toward the NN. 
    The reciprocal space coordinates of the peaks are:
    $(02)$ is at 
    $Q_\parallel = 1.547$~\AA{}$^{-1}$, 
    $Q_z = 0$~\AA{}$^{-1}$ and 
    ($11$), ($1\overline 1$) are at 
    $Q_\parallel = 1.491$~\AA{}$^{-1}$, 
    $Q_z = 0.3$~\AA{}$^{-1}$.
    From these coordinates one can directly infer the parameters of the elementary unit of a 2D lattice: 
    $a_1 = 4.93\pm0.02$~\AA{}, 
    $a_2 = 8.12\pm0.02$~\AA{} 
    with the lattice angle $\gamma = 90^\circ$.
    The GID pattern in Fig.~\ref{fig:SRO}(e) corresponds to the monolayer in its final state, i.e., after structural rearrangement.
    Only the broadened peak at 
    $Q_\parallel = 1.547$~\AA{}$^{-1}$, 
    $Q_z = 0$~\AA{}$^{-1}$ corresponding to short-range positional order presents, 
    indicating lateral disordering of the monolayer.

    The initial state of the third sample,
    measured just after monolayer compression,
    corresponds to a polycrystalline solid-state phase. 
    Such state was previously described using the QLRO model, 
    it can also be effectively analyzed using the SRO model, 
    provided that the degree of disorder is relatively weak.
    The parameters for GID simulation using SRO model in Fig.~\ref{fig:SRO}(b) are chosen to agree with the experiment 
    [cf. Fig.~\ref{fig:SRO}(a)].
    The GID simulations in Figs.~\ref{fig:SRO}(b)--\ref{fig:SRO}(d) demonstrates gradual transition from polycrystalline solid-state phase to paracrystalline phase with short-range correlations. 
    The simulations are performed using SRO model with the same lattice parameters, 
    but with different $\sigma_z$ and different order parameter $s = \sigma/a$, 
    which  represents the standard deviation in terms of fractions of the lattice parameter (see section \ref{sec:SRO}). 
    This normalization enables an equivalent treatment of in-plane disorder along each basis vector.
    The value of parameter $s$ is given in percentage.
    This parameter can be used to estimate the correlation length associated with the region over which positional correlations between molecules remain.
    Note again that the parameters for GID simulation in Fig.~\ref{fig:SRO}(d) are chosen to agree with the experiment in Fig.~\ref{fig:SRO}(e).
    In this case, 
    the width of the Bragg peaks in the $Q_\parallel$ direction is determined not only by $M$, but also by $s$. 
    The greater the dispersion of molecular deviations related to the parameter $s$,
    the shorter the correlation length, and the broader the peak.
    Note, that the SRO model more accurately describes the integral distribution of intensity between peaks in the GID pattern. 
    It is important, 
    that SRO model does exhibit the peak broadening with increase of parameter $s$, 
    which is in contrast to QLRO model,
    where such broadening is absent~\cite{de2003structure}.
    
    Thus, the first monolayer sample compressed to a surface pressure of 20~mN/m at a temperature of 23$^\circ$C exhibits a standard diffraction pattern 
    that corresponds to the polycrystalline solid-state phase of the arachidic acid monolayer. 
    The interaction with Ce$^{3+}$ ions has affected the arrangement of molecules in the monolayer, 
    the lattice is not hexagonal ($a_2 \neq \sqrt{3} a_1$), 
    but the lattice angle remains the same, $\gamma = 90^\circ$.
    When the monolayer samples are compressed to a surface pressure beyond the collapse point,
    there are two possible reorganization scenarios.
    A typical case of collapse is shown in the third monolayer sample compressed to a surface pressure beyond the collapse of 52~mN/m at a temperature of 23$^\circ$C. 
    Note that, according to BAM data~\cite{nikolaev2025probing} taken at $T = 23^\circ$C,
    the monolayer exhibits the formation of large 3D crystalline-like aggregates. 
    Some part of the monolayer remains at the liquid-air interface and reorganizes into a paracrystal.
    The GID data in Fig.~\ref{fig:SRO} clearly show the process of paracrystalline formation in the collapse state monolayer characterized by SRO.
    Remarkably, a different GID pattern [Fig.~\ref{fig:slope}(d)] was recorded for the second sample compressed to a surface pressure beyond the collapse point at the low temperature ($T=21^\circ$C).
    Rather complicated GID patterns for this sample reveal challenging collapse events that unfold in Langmuir monolayer with PHB packing mode. 
    Specifically, the structural rearrangement in overcompressed PHB monolayer can be attributed to the corrugation of individual crystallites,
    whereas molecules in each crystallite keep high degree of structural ordering. 
    This intricate collapse behavior can be explained by the characteristic mutual orientation of neighboring aliphatic chains,
    observed in monolayer at low temperature,
    when molecular rotational energy diminishes. 
    The enhanced ordering enables a further decrease of area per molecule, 
    resulting in the formation of an even denser monolayer.

\section{Conclusion}

    In this study, a theoretical framework for simulating GID from Langmuir monolayers is developed. 
    This approach is based on the DWBA, which treats the monolayer as a perturbation of a planar air-liquid interface.
    This formalism accounts for the diffraction and scattering of evanescent waves along the monolayer surface.
    While the application of DWBA to scattering in Langmuir monolayers is well established, 
    we have extended the theory to accommodate a broader range of in-plane structural changes.
    The various types of lateral order now can be straightforwardly plugged into the calculation.
    The only remaining task when dealing with a new type of ordering is to derive the corresponding correlation function.
    This generalization enhances the flexibility of the model and facilitates its application to systems undergoing  structural reorganization within the monolayer.

    To demonstrate the versatility of the proposed approach, 
    we examined the case of the structural reorganization of an arachidic acid monolayer interacting with cerium (Ce$^{3+}$) ions.
    In its solid-state phase, 
    the monolayer exhibits a typical two-dimensional polycrystalline structure characterized by NN tilt of the aliphatic chains.
    However, the structure of this system can vary significantly in a collapsed state.
    At sufficiently high temperature and surface pressure beyond the collapse point, 
    the monolayer loses its crystallinity, 
    which is followed by the formation of three-dimensional aggregates on the surface.
    The remnants of the monolayer exhibit paracrystalline organization as evidenced by the GID.
    However, at lower temperatures and pressures beyond the collapse point, the monolayer retains its polycrystalline structure. 
    In this case, mechanical stress induced within the system is released through the corrugation of the monolayer.

    We applied our theoretical framework to simulate all three structural states observed in the system.
    For the polycrystalline solid-state phase, the GID pattern was simulated using the well-known QLRO model. 
    The resulting two-dimensional diffraction patterns revealed that the actual molecular tilt is governed not only by the position of the GID peak along the $Q_z$-axis, 
    but also significantly influenced by the parameter $\sigma_z$, 
    which characterizes the out-of-plane disorder.
    To describe the gradual transition from the polycrystalline solid-state phase to a paracrystalline organization at elevated temperatures,
    we employed the SRO model. 
    By varying the order parameter, 
    we demonstrated the broadening of the diffraction peak, 
    indicative of the reduced length of positional correlations. 
    The most compelling results were obtained in the collapsed state of the monolayer at lower temperatures. 
    To describe a new type of ordering,
    specifically the monolayer corrugation, 
    we introduced the model of mosaicity around the vertical direction.
    This flexible framework enables accurate quantitative analysis of two-dimensional diffraction maps obtained from Langmuir monolayers via grazing-incidence diffraction.
    Overall, our quantitative analysis provides new insights into the complex structural dynamics of soft matter monolayers.
    These findings underscore the importance of a versatile approach to simulations of full two-dimensional GID patterns in order to elucidate the diverse ordering mechanisms present in Langmuir monolayers.

\section*{Acknowledgments}

    The authors acknowledge the scientific infrastructure of the Kurchatov Synchrotron Radiation Source "KISI" at the NRC "Kurchatov Institute", 
    which made the synchrotron measurements possible.
    
\section*{Funding information}
    
    This research was funded by the Ministry of Science and Higher Education of the Russian Federation (grant no. FSFZ-2024-0003). 
    The participation of BIO in the theoretical part of this research was supported by Russian Science Foundation (grant no. 23-12-00200).

\bibliographystyle{ieeetr} 
\bibliography{references}
 
\end{document}